\documentclass[conference]{IEEEtran}
\IEEEoverridecommandlockouts
\usepackage{cite}
\usepackage{amsmath,amssymb,amsfonts}
\usepackage{algorithmic}
\usepackage{graphicx}
\usepackage{textcomp}
\usepackage{xcolor}

\usepackage{algorithm}
\usepackage[hyphens]{url}
\usepackage{caption}
\usepackage{hyperref}

\def\BibTeX{{\rm B\kern-.05em{\sc i\kern-.025em b}\kern-.08em
    T\kern-.1667em\lower.7ex\hbox{E}\kern-.125emX}}
\begin{document}

\title{Contextual Knowledge Sharing in Multi-Agent Reinforcement Learning with Decentralized Communication and Coordination}

\author{\IEEEauthorblockN{Hung Du, Srikanth Thudumu, Hy Nguyen, Rajesh Vasa, Kon Mouzakis}
\IEEEauthorblockA{
\textit{Applied Artificial Intelligence Institute ($A^2I^2$),  Deakin University} \\
Geelong VIC 3216, Australia  \\
Emails: \{hung.du, srikanth.thudumu, hy.nguyen, rajesh.vasa, kon.mouzakis\}@deakin.edu.au}
}

\maketitle

\begin{abstract}
Decentralized Multi-Agent Reinforcement Learning (Dec-MARL) has emerged as a pivotal approach for addressing complex tasks in dynamic environments. Existing Multi-Agent Reinforcement Learning (MARL) methodologies typically assume a shared objective among agents and rely on centralized control. However, many real-world scenarios feature agents with individual goals and limited observability of other agents, complicating coordination and hindering adaptability. Existing Dec-MARL strategies prioritize either communication or coordination, lacking an integrated approach that leverages both. This paper presents a novel Dec-MARL framework that integrates peer-to-peer communication and coordination, incorporating goal-awareness and time-awareness into the agents' knowledge-sharing processes. Our framework equips agents with the ability to (i) share contextually relevant knowledge to assist other agents, and (ii) reason based on information acquired from multiple agents, while considering their own goals and the temporal context of prior knowledge. We evaluate our approach through several complex multi-agent tasks in environments with dynamically appearing obstacles. Our work demonstrates that incorporating goal-aware and time-aware knowledge sharing significantly enhances overall performance.
\end{abstract}


\section{Introduction}

\begin{figure*}
    \centering
    \includegraphics[width=0.8\textwidth,height=0.5\textwidth]{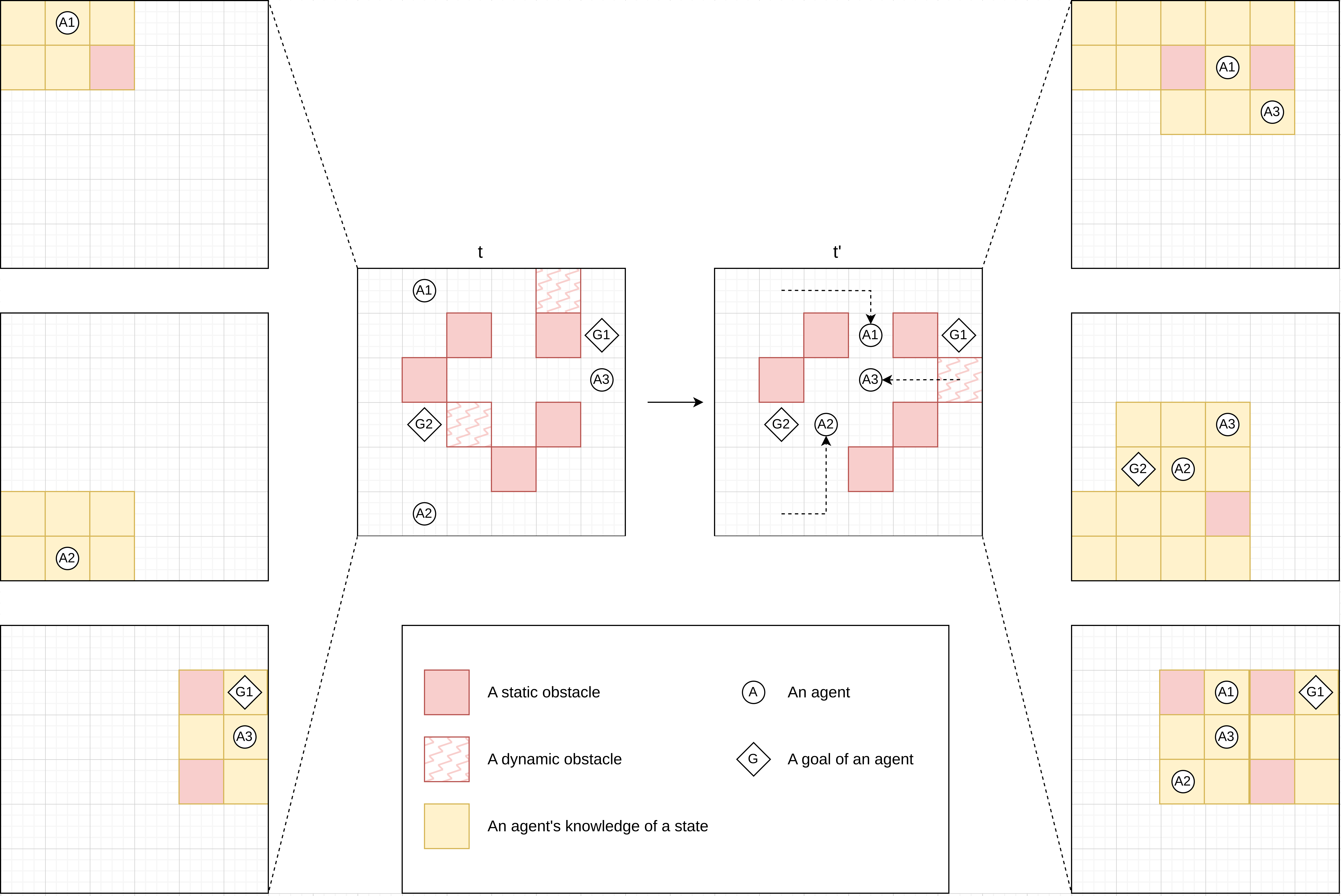}
    \caption{An illustration of a fully decentralized environment with multiple agents ($t < t'$). While the goal of Agents 1 and 2 is G1, that of Agent 3 is G2. Note that at time $t'$, Agent A3 is unaware of an obstacle that has occurred in a position it previously encountered, rendering its knowledge about that location obsolete. Additionally, during a communication session with A3, Agents A1 and A2 must be aware of the outdated information provided by A3 to select the optimal action.}
    \label{fig:example1}
\end{figure*}

Cooperative Multi-Agent Reinforcement Learning (MARL) has emerged as a critical research area due to its potential to overcome the limitations of single-agent systems in addressing complex, real-world problems. While single-agent systems have demonstrated success in achieving human-like performance in specific scenarios \cite{du2024survey}, they often face limitations in terms of scalability, adaptability, and reliability, especially when dealing with complex tasks that require specialized agents \cite{kantamneni2015survey, de2019survey, amirkhani2022consensus}. To address these limitations, the multi-agent system (MAS) architecture has gained prominence, enabling agents to communicate, coordinate, and tackle complex tasks in dynamic environments. MARL plays a key role in handling such dynamics \cite{gronauer2022multi,du2024survey}. Among the various approaches within MARL, the Centralized Training and Decentralized Execution (CTDE) paradigm \cite{kraemer2016multi,gupta2017cooperative} is popular for cooperative tasks \cite{lowe2017multi,foerster2018counterfactual,wang2020shapley,rashid2020monotonic,wang2020roma,ruan2022gcs,nayak2023scalable}. This approach employs a centralized critic during training to develop decentralized policies for agents, which are then executed independently. Although widely adopted, CTDE-based algorithms encounter significant difficulties in environments with large joint state-action spaces and inherent stochasticity. Moreover, these algorithms typically assume that agents share a common goal and depend on centralized control. However, many real-world situations involve agents with individual objectives and limited observability of others, leading to potential miscoordination, sub-optimal policies, and reduced adaptability.

The Decentralized Training and Decentralized Execution (DTDE) paradigm \cite{tan1993multi,jiang2022i2q} aims to address the limitations of the CTDE approach by relaxing the assumptions of full observability and centralized control. In a fully decentralized setting, each agent operates with its own goals and observations, communicates with other agents within its observation range, and coordinates during these communication sessions. The agent then uses the acquired observations and knowledge to optimize its objectives. This approach has the potential to enhance the robustness and adaptability of agents in handling uncertainties. However, DTDE-based algorithms can face significant challenges, including (a) exhaustive exploration due to the absence of a centralized coordinator and limited observability, and (b) inefficient sharing of experience and knowledge caused by the growing number of agents and the rapid obsolescence of information.

A promising approach to reducing the exhaustive exploration of independent agents in DTDE-based algorithms is to establish a communication protocol among agents. A straightforward communication scheme allows agents to share their local observations, which can then be used to optimize their local policies toward individual goals \cite{singh2018learning,jiang2018learning,das2019tarmac,zhang2019efficient,jiang2022i2q}. While this approach reduces exploration time, it also introduces a significant amount of irrelevant information, which increases learning complexity and can degrade performance. Several strategies have been proposed to address this challenge. For example, agents can be instructed on when to communicate \cite{liu2020who2com,liu2020when2com,hu2022where2comm}. In team settings, incentive-based communication schemes have been used to filter out trivial information and promote coordination toward a global objective \cite{yuan2022multi}. Additionally, integrating Graph Neural Networks (GNNs) with agent feature embeddings and mutual information has been applied to eliminate irrelevant information \cite{ding2024learning}. Other methods focus on pruning irrelevant agents from communication sessions by leveraging agents' identities \cite{du2024expressive} and personalized communication topology \cite{meng2024pmac}. Furthermore, approaches to address communication bandwidth limitations have been introduced by advising mechanisms \cite{da2017simultaneously,ba2024cautiously} and message pruning \cite{ding2020learning,mao2020learning}. Although these strategies show promise in addressing communication challenges among agents, they often assume that decentralized agents share the same local objective. However, even within the same team or coalition, agents may pursue different individual goals (see also Figure \ref{fig:example1}). Therefore, goal awareness becomes essential to improve the effectiveness of communication among agents.

Coordination strategies facilitate the efficient sharing of experience and knowledge among independent agents in DTDE-based algorithms. One widely-used strategy involves utilizing a global value, estimated by aggregating the local values of states and actions across agents \cite{sunehag2018value,rashid2020monotonic}. Furthermore, graph-based approaches \cite{bohmer2020deep,ruan2022gcs,pesce2023learning,nayak2023scalable} have been employed to represent the relationships between observations or agents, which are then used to enhance agent coordination. Advising mechanisms \cite{da2017simultaneously,ba2024cautiously} also play a crucial role by encouraging experienced agents to offer guidance to less experienced agents, based on their knowledge. These mechanisms further motivate agents to explore novel states in the environment, which can be also achieved by estimating intrinsic rewards through weighted mutual information between agents' novel states \cite{jiang2024settling}. However, these strategies assume that observations and knowledge remain constant over time. In practical scenarios, the value of information decays and eventually becomes invalid, leading to sub-optimal policies. Therefore, time awareness is essential for improving the effectiveness of coordination among agents.

In this paper, we propose a novel Dec-MARL framework designed to address two key challenges in fully decentralized settings: exhaustive exploration and inefficient knowledge sharing among agents. Our framework integrates peer-to-peer communication and coordination, incorporating both goal awareness and time awareness to provide agents with two primary capabilities. First, goal-aware communication enables agents to exclude irrelevant agents during communication sessions. Second, agents can retrieve relevant observations and share their knowledge by understanding the goals of other agents. Additionally, we introduce a time factor that decays the value of information over time, along with a novel intrinsic reward mechanism that encourages agents to explore new states in the environment. We evaluate our framework using complex multi-agent tasks in a grid world environment where obstacles dynamically appear. Our experiments demonstrate that our framework enhances agents' exploration and knowledge sharing in fully decentralized environments.

\section{Related Work}

\subsubsection{Decentralized Training and Decentralized Execution (DTDE)}
Approaches in cooperative MARL typically fall into two categories: Centralized Training and Decentralized Execution (CTDE) \cite{kraemer2016multi, gupta2017cooperative} and DTDE \cite{tan1993multi}. CTDE-based methods \cite{lowe2017multi,foerster2018counterfactual,wang2020shapley,rashid2020monotonic,wang2020roma,ruan2022gcs,nayak2023scalable} have demonstrated the stability of training multiple agents for cooperative tasks in complex environments. These approaches often assume that agents have unlimited access to all states in the environment and rely on centralized control for assessing agents' actions. However, such assumptions are not feasible in many real-world scenarios where environments are dynamic, agents have limited observability, and may pursue different individual goals. As a result, the robustness of CTDE-based approaches diminishes in these situations. Conversely, DTDE-based methods \cite{tan1993multi,de2020independent,jiang2022i2q,jin2022v,daskalakis2023complexity,skrynnik2024learn} do not require full observability or centralized control among agents. Despite this, DTDE approaches often encounter challenges such as exhaustive exploration and inefficient knowledge sharing, due to several factors: (a) the absence of efficient communication and coordination strategies; (b) the lack of centralized control; (c) the increasing number of agents; and (d) the rapid changes in information within dynamic environments. To address these challenges, we propose a novel DTDE-based framework that equips agents with goal awareness and time awareness and integrates communication and coordination among agents in fully decentralized settings.

\subsubsection{Communication}
Communication is vital in overcoming the challenge of exhaustive exploration in MARL approaches. A naive design involves establishing a communication protocol among all agents within the same environment \cite{singh2018learning,jiang2018learning,das2019tarmac,zhang2019efficient}. However, this approach can hinder agent performance due to the curse of dimensionality, as the number of agents increases and information overload becomes an issue. Existing methods that aim to mitigate this challenge can be categorized into three groups: (a) communication-triggering instructions \cite{liu2020who2com,liu2020when2com,hu2022where2comm}; (b) filtering out irrelevant information \cite{yuan2022multi,ding2020learning,mao2020learning,ding2024learning}; and (c) filtering out irrelevant agents \cite{meng2024pmac,du2024expressive}. These methods typically operate within the CTDE framework and assume that agents share the same local objectives. Our framework differs from these approaches in two key ways: (i) agents operate in a fully decentralized setting with limited communication; and (ii) agents are equipped with goal awareness, enabling them to understand the goals of other agents before initiating communication sessions.

\subsubsection{Coordination}
Coordination strategies are crucial for effective knowledge sharing among agents. Existing approaches often involve aggregating information among agents \cite{sunehag2018value,rashid2020monotonic,christianos2020shared} and improving this process by also considering the relationships between pieces of information \cite{bohmer2020deep,ruan2022gcs,pesce2023learning,nayak2023scalable}. These strategies, however, typically operate within the CTDE framework, which can hinder coordination in many real-world scenarios. Another form of coordination involves motivating agents to explore novel states in the environment through advice \cite{da2017simultaneously,ba2024cautiously} or intrinsic rewards \cite{devidze2022exploration,jiang2024settling}. However, these approaches often overlook the fact that the value of information decays over time and can become obsolete, leading to inefficient knowledge sharing. Our framework addresses this by incorporating both time awareness and goal awareness to enhance agent coordination. We also introduce a novel reward function that uses a time-aware intrinsic reward to motivate agents to explore new states and revisit previously known states to refresh their knowledge.

\section{Problem Preliminary}
% MARL
We formulate our framework as the Decentralized Multi-Agent Reinforcement Learning (Dec-MARL) where the decision-making process of an agent follows Partially Observable Markov Decision Process (POMDP) \cite{oliehoek2016concise}. This is defined as follows: $(n, \mathcal{S}, \{\mathcal{A}_i\}_{i=1}^{n}, T, \{\mathcal{R}_i\}_{i=1}^{n}, \{\mathcal{O}_i\}_{i=1}^{n}, P, \gamma)$ where $n$ is the number of agents, $\mathcal{S}$ is the set of states, $\{\mathcal{A}_i\}_{i=1}^{n}$ denotes the set of action sets for each agent, $T: \mathcal{S} \times \mathcal{A}^n \rightarrow \mathcal{S}'$ is the state transition probability function following the joint actions $\mathcal{A}^n = (a_{1}, a_{2}, \ldots, a_{n})$, $\{\mathcal{R}_i\}_{i=1}^{n}$ is the set of rewards for each agent, $\{\mathcal{O}_i\}_{i=1}^{n}$ represents the set of observations for each agent, $P: \mathcal{S} \times \mathcal{A}^n \rightarrow \mathcal{O}'$ is the observation probability function, and $\gamma \in [0, 1]$ is the discount factor. Furthermore, a set of individual goals of agents is denoted by $\{\mathcal{G}_{i}\}_{i=1}^{n}$ where goals can be defined in terms of states $\mathcal{G} \subseteq \mathcal{S}$ \cite{schaul2015universal}. Notably, an agent can only access its own local observations and learn an independent policy $\pi_{i}$ to maximize its own goal in the decentralized setting.

% Time and Mental State
In a fully decentralized environment, agents' observations can be limited and vary from one another (see also Figure \ref{fig:example1}). As a result, an agent only possesses knowledge of the states it has experienced, leaving other states unknown. Moreover, the value of acquired knowledge diminishes over time due to the environment's dynamics. This necessitates that agents consider the time factor associated with such knowledge when adjusting their policies. In our framework, we model this behavior by introducing a mental state for each agent, denoted by $\{\mathcal{M}_{i}\}^{n}_{i=1}$. It is important to note that, within our framework, all agents share the same ontology and bounded environment. Consequently, the mental state of an agent encompasses all masked states in the environment, defined as follows: $\mathcal{M}_{i} = \{(s, m_{t}, d_{t}) \}_{s\in \mathcal{S}}$, where $m$ represents the masked label of state $s$ at time step $t$ (e.g., \textit{empty}, \textit{obstacle}, \textit{unknown}, or other), and $d_{t}$ denotes the duration since the last visit. Additionally, $m_{t}$ dynamically changes in response to the environment.

\section{Method}
In this section, we propose a novel Dec-MARL framework that equips agents with goal awareness and time awareness for addressing challenges of communication and coordination in full decentralized environments. Furthermore, detailed explanations of each component of our framework are provided below.

\subsection{Representations and Value Approximation}

\begin{figure}
    \centering
    \includegraphics[width=0.7\linewidth,height=0.65\linewidth]{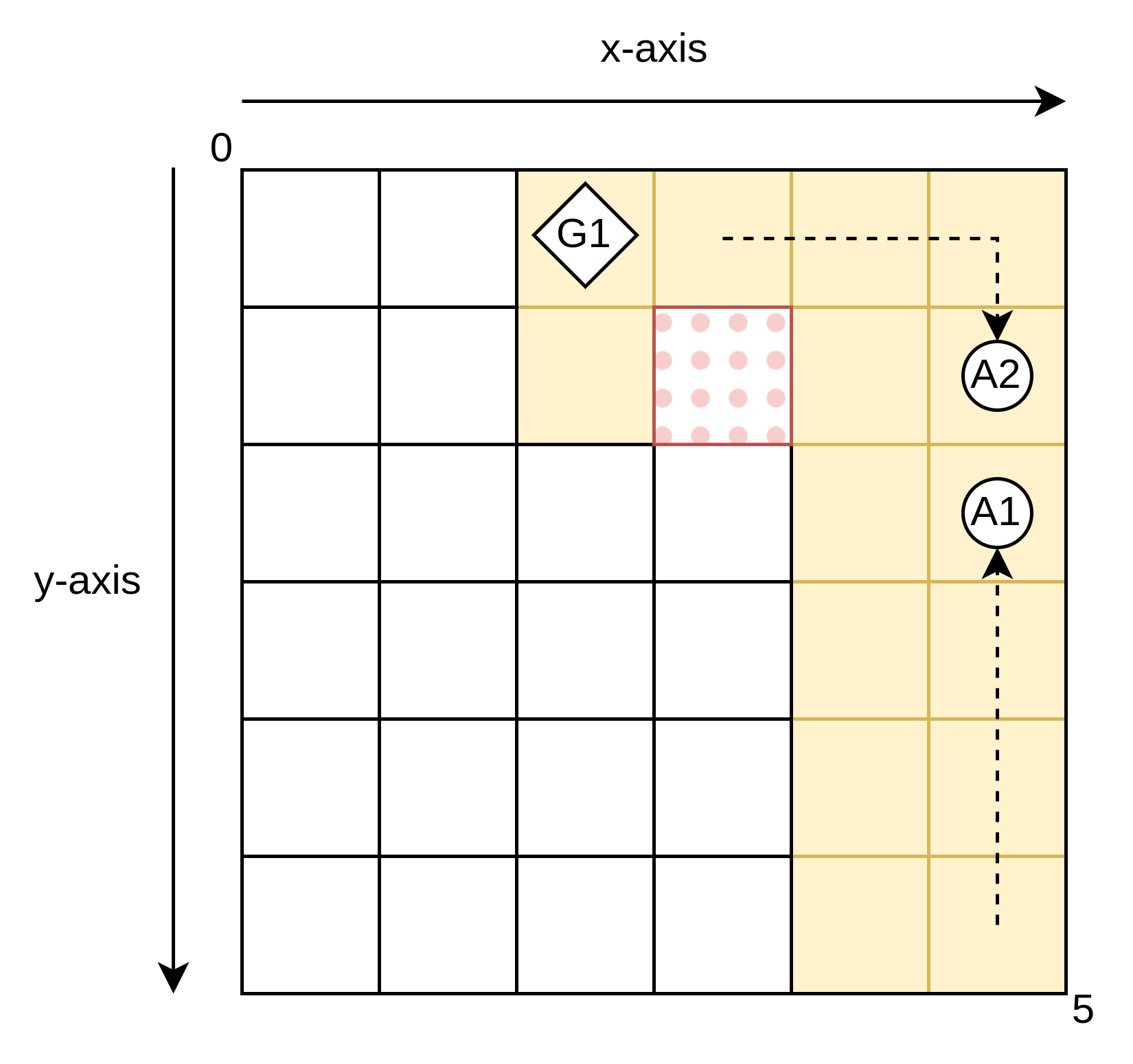}
    \caption{A demonstration of the intrinsic reward guiding Agent 1 (A1) to choose an action that optimizes both the goal-oriented objective and the exploration of uncertainty. The filled yellow boxes represent knowledge of A1 in terms of that position, the red box filled by dots is obstacle, and the remaining are unknown to A1. In this scenario, the optimal action for A1 is to move towards the $(4, 2)$ position, as it strikes a balance between both objectives.}
    \label{fig:example2}
\end{figure}

To facilitate generalization, our framework encodes the following properties of the agent: ($s$, $g$, $o$, $m$, $a$). Specifically, we define $f_{x}(x) \rightarrow \mathbf{e}_{x} \in \mathbb{R}^{k}$ as the representation function, which could involve methods such as one-hot encoding, Multi-Layer Perceptron (MLP), categorical encoding, image-based encoding, or other representation techniques. Here, $x$ represents one of the agent's properties, and $k$ denotes the dimensionality of the embedding. It is important to note that both $f$ and $k$ can vary depending on the specific agent property. Furthermore, the mental state of an agent is represented as follows:
\begin{equation} \label{eq:ems}
    \mathbf{e}_{\mathcal{M}} = \bigcup_{(s, m) \in \mathcal{M}} \left( \mathbf{e}_s \oplus \mathbf{e}_m \right)
\end{equation}
where $\oplus$ is the concatenation operation between two embeddings between $\mathbf{e}_s$ and $\mathbf{e}_m$, and $\bigcup$ is the aggregation function (e.g., summation, dot product, average pooling, or other). Our framework applies the average pooling. Note that the scheme of Equation \ref{eq:ems} excludes the time factor.

Understanding its current goal and recent mental state is essential for an agent to adjust its actions in two key ways: (a) moving toward the goal based on its belief about future states, or (b) exploring uncertain states that could be advantageous for achieving the current goal. As illustrated in Figure \ref{fig:example2}, there are situations where the agent must balance these two aspects to maximize rewards. Inspired by the Universal Value Function Approximator (UVFA) \cite{schaul2015universal}, our framework integrates both the agent's goal and mental state into the construction of the policy function as $\pi: \mathcal{S} \times \mathcal{G} \times \mathcal{M} \rightarrow \mathcal{A}$. The corresponding action-value function is then defined as $Q(s, a, g, \mathcal{M}; \theta^{Q}) \approx Q^{*}_{g, \mathcal{M}}(s, a)$ where $\theta^{Q}$ is learning parameters.

\subsection{Time Awareness and Intrinsic Rewards}
\begin{figure}
    \centering
    \includegraphics[width=\linewidth,height=0.65\linewidth]{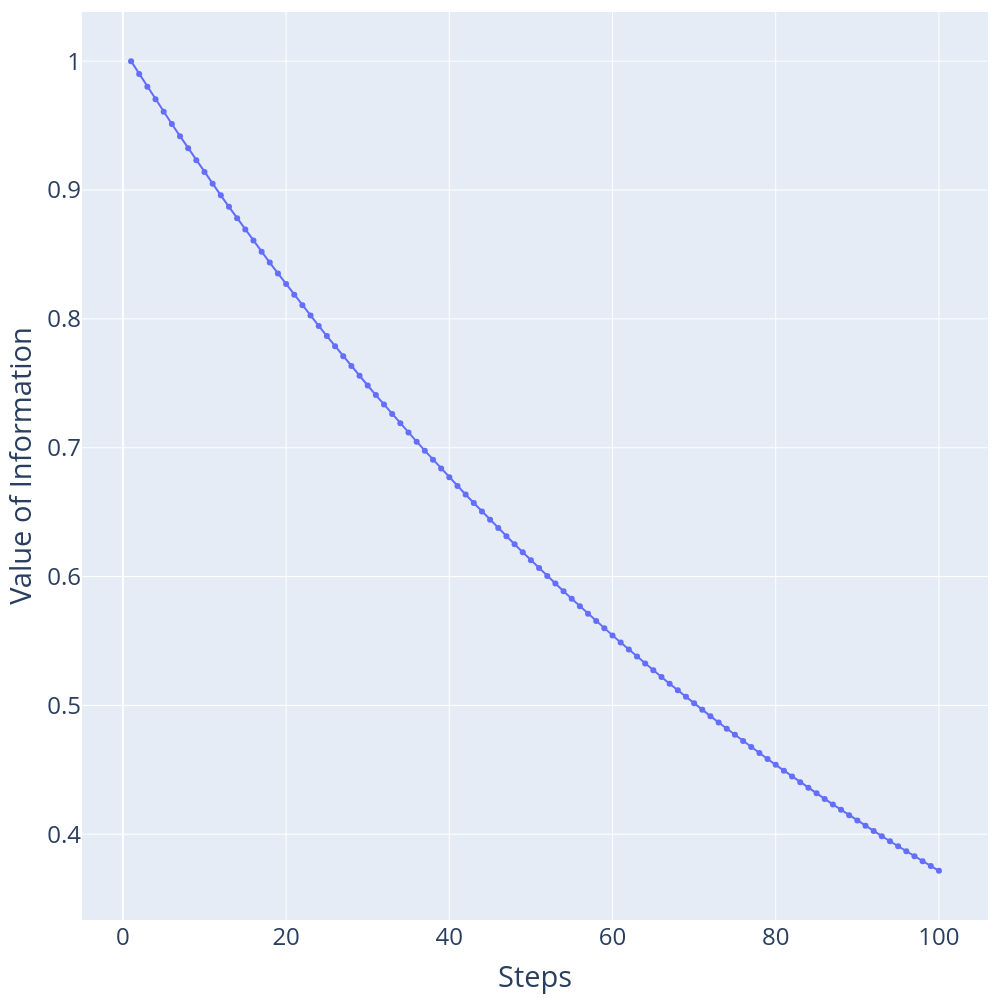}
    \caption{An illustration of utilizing Equation \ref{eq:time-nov} to estimate the novelty of information over 100 steps where $d_{t'}$ is estimated with the time increment of 0.01 as: $t' = t + 0.01$. Importantly, in this graph, we assume that the information is not reflected by an agent per step.} 
    \label{fig:example3}
\end{figure}

A reward provided by the environment is designed to guide an agent toward achieving its goal, commonly referred to as an extrinsic reward. In decentralized training, the agent does not have access to the global state. Therefore, exploring novel observations that are based on the agent's local observations and potentially beneficial for future outcomes can be encouraged by using an intrinsic reward. The novelty of an observation is often estimated using a utility function with count-based mechanisms \cite{jiang2024settling} as follows:
\begin{equation} \label{eq:count-nov}
    u^{t}_{i}(o) = \frac{1}{\mathcal{N}_{o}}
\end{equation}
where $u^{t}_{i}$ is dependent on the local observations of agent $i$ at time $t$, and $\mathcal{N}_{o}$ is the frequency of the observation. Additionally, $u^{t}_{i}$ can vary between agents. Equation \ref{eq:count-nov} indicates that the novelty of an observation decreases as it occurs more frequently in the agent's experience. However, in many practical scenarios, an observation may become novel again despite its high frequency. This is due to the dynamics of the environment. Hence, instead of using count-based mechanisms, we introduce the time factor that measures the novelty of an observation as:
\begin{equation} \label{eq:time-nov}
    u^t_{i}(o) = e^{\frac{1}{2} d_{t'}}
\end{equation}
where $e$ is the exponential function, $d_{t'} \in \mathcal{M}_{i}$, and $t' \leq t$. In addition, $d_{t'}$ is controllable according to the application domains. From our empirical analysis, we would suggest keeping $d_{t'}$ as small as possible with the time increment less than 0.1 in situations where information is gradually changed (see also Figure \ref{fig:example3}). Equation \ref{eq:time-nov} satisfies the following two conditions: (a) the value of the observation decays over time after being uncovered by the agent; and (b) the observation becomes novel again after being re-discovered by the agent. Furthermore, the value of $u^{t}_{i}$ can be integrated with embeddings in Equation \ref{eq:ems} and convert $\mathbf{e}_{\mathcal{M}}$ into the time-aware scheme as follows:
\begin{equation} \label{eq:ems_t}
    \mathbf{e}^{t}_{\mathcal{M}} = \bigcup_{(s, m) \in \mathcal{M}} \left( u^{t}_{i}(s) \cdot \left( \mathbf{e}_s \oplus \mathbf{e}_m \right) \right)
\end{equation}
where $u^{t}_{i}(s) \in \mathbb{R}$ is a scalar value.

In addition to being integrated with the embedding of the mental state of an agent, we introduce a novel reward estimation that combines both the extrinsic reward and $u^{t}_{i}$ as the intrinsic reward as follows:
\begin{equation} \label{eq:reward}
    r^{s}_{i} = (1 - \alpha) r_{\text{ext}} + \alpha \frac{1}{|\mathcal{M}|}\sum_{s' \in \mathcal{M}}{u^{t}_{i}(s')}
\end{equation}
where $r_{\text{ext}}$ is the extrinsic reward of the agent and can be customized in terms of application domains, $|\mathcal{M}|$ is the number of states in the agent's mental state, and $\alpha \in [0, 1]$ is the dampening factor that balances two types of rewards. Moreover, as shown in Equation \ref{eq:reward}, the intrinsic reward increases when the agent continues to explore new states or revisits old ones. However, the agent is not solely biased toward exploration; instead, it aims to move toward states that balance both factors.

\subsection{Integration of Communication and Coordination}

In fully decentralized settings, it is essential for agents to communicate and share relevant observations and knowledge. While relevant observations can accelerate an individual agent's exploration, relevant knowledge can enhance their performance in achieving their goals. However, shared information can have both positive and negative impacts on an agent's policy and action value function \cite{ba2024cautiously,ding2024learning}. Therefore, it is important for agents to carefully evaluate the information they receive before incorporating it into their current policy and action value function. In our framework, we propose a strategy that integrates communication and coordination, incorporating goal awareness and time awareness. This strategy consists of three phases: Share-Reason-Aggregate. The details of each phase are specified below.

\subsubsection{Share}
As the agent navigates the environment, it may encounter other agents within its observation range, allowing for the establishment of communication and coordination sessions. During a communication session, the agent broadcasts its goal to identify two types of agents: (a) agents who share the same goal, known as current peers, and (b) agents who have relevant knowledge from their experience but do not share the same goal, referred to as current advisors. It is worth noting that peers do not necessarily have prior experience of the given goal. Once this identification process is complete, the agents initiate the coordination session. Both peers and advisors retrieve observations relevant to the goal. The retrieval mechanisms can differ depending on the problem domain. In our framework, a peer retrieves both observations from its mental state and learning parameters such as $\theta^{\pi}$ and $\theta^{Q}$. Additionally, inspired by \cite{skrynnik2024learn}, each agent in our framework is equipped with a heuristic planning capability that is activated only when the agent is in the role of an advisor. Specifically, in discrete observation spaces, such as a 2D map $(x, y)$, an advisor estimates the shortest path comprising observations between the agent’s current position and the given goal. Advisors do not share their learning parameters, as these parameters are optimized for different goals that may not align with the agent's current goal. It is important to note that agents in our framework share the same ontology and bounded environment, making observations and knowledge transferable among them.

\subsubsection{Reason}
After the knowledge-sharing process, the agent activates its reasoning capability rather than blindly following the acquired observations and knowledge. To achieve this, our framework equips agents with a rule-based reasoning capability. First, the agent reflects on its mental state using the latest and novel observations shared by peers and advisors as follows:
\begin{equation}
    \mathcal{M}_{i} = \bigcup_{j=1; s \in \mathcal{S}}^{K} \begin{cases}
        (s, m_{t}, d_{t})_{i}, & \text{if } \left(d_t\right)_i   <  \left(d_t\right)_j \\
        (s, m_{t}, d_{t})_{j}, & \text{if } \left(d_t\right)_i   >  \left(d_t\right)_j \lor (s)_{j} \notin \mathcal{M}_{i} \\
    \end{cases}
\end{equation}
where $\bigcup$ is the set union function, $K$ is the total number of peers and advisors and $j$ represents the index of a peer or an advisor. Second, to determine whether to update its learning parameters, the agent estimates the overlap ratio between its mental state and the observations shared by each peer. In our framework, this overlap ratio between discrete observations is calculated using the Jaccard similarity as:
\begin{equation}\label{eq:jac}
    J(\mathcal{M}_{i}, \mathcal{M}_{j}) = \frac{
        |\{ (s, m_{t}) \}_{i} \cap \{ (s, m_{t}) \}_{j}|
    }{
        |\{ (s, m_{t}) \}_{i} \cup \{ (s, m_{t}) \}_{j}|
    } 
\end{equation}
where $J \in [0, 1]$, with $J = 0$ indicating that agents $i$ and $j$ have no overlapping observations, and $J = 1$ indicating a complete match between their mental states. Here, $s$ represents the known state of an agent. It is important to note that this estimation takes place before agents update their mental states with the newly acquired observations. The primary objective is to encourage the agent to integrate novel knowledge obtained from its peers.

\subsubsection{Aggregate}
After selecting peers based on the overlap ratio, the agent updates its learning parameters as follows:
\begin{equation} \label{eq:agg_policy}
    \theta^{\pi}_{i} = (1 - \beta) \theta^{\pi}_{i} + \beta \frac{1}{K}\sum_{j=1}^{K}{\theta^{\pi}_{j}}
\end{equation}
\begin{equation} \label{eq:agg_value}
    \theta^{Q}_{i} = (1 - \beta) \theta^{Q}_{i} + \beta \frac{1}{K}\sum_{j=1}^{K}{\theta^{Q}_{j}}
\end{equation}
where $K$ represents the total number of selected peers, and $\beta$ is the dampening factor that balances the agent's own learning parameters with those aggregated from its peers. Our empirical analysis indicates that Equations \ref{eq:agg_policy} and \ref{eq:agg_value} can sometimes lead to situations where poor-performing agents negatively impact the performance of others during the coordination session. Therefore, we recommend keeping $\beta$ as low as possible.

\section{Experiments}
\subsection{Environments and Tasks}
To evaluate our framework, we designed a 2D map with dynamically appearing obstacles. The environment comes in two sizes: Base (10 x 10) and Large (20 x 20), to test the scalability of our framework. Each environment features 3 objects surrounded by obstacles, with the number and positions of these obstacles being static. We created two difficulty levels for the environment: Easy and Hard. In the Easy environment, obstacles remain unchanged over time, while in the Hard environment, obstacles can appear and disappear dynamically. Specifically, in the Hard environment, an obstacle may appear at time $t$ and disappear at a later time $t'$, where $t' > t$, or vice versa. The Hard environment is designed to assess how well agents in our framework handle environmental dynamics.

Combining the environment sizes with the difficulty levels results in four distinct environments: Base-Easy, Base-Hard, Large-Easy, and Large-Hard. In each environment, an agent starts at a predefined position far from its goal, with agents being placed in different areas far from one another. There are two scenarios regarding the agents' goals: (i) all agents pursue the same goal, and (ii) there are two distinct goals, with at least two agents pursuing the first goal and the remaining agents pursuing the second goal. Moreover, multiple agents can occupy the same cell. An agent's task is considered complete if it reaches its goal and remains in that position.

\subsection{Implementation Details}
We conducted our experiments a complex 2D environment with fully decentralized settings. Here, $s^{t}_{i} = (x^{t}_{i}, y^{t}_{i})$, $\mathcal{G} \subset \mathcal{S}$, $m$ can be one of the following labels: \textit{empty}, \textit{obstacle}, \textit{object}, \textit{agent}, or \textit{unknown}, and $a$ can be one of the following options: \textit{left}, \textit{right}, \textit{up}, \textit{down}, or \textit{stay}. Furthermore, we utilized categorical encoding functions to represent agent's properties ($s$, $g$, $o$, $m$, $a$) in their embeddings as: $\mathbf{e}_{s} \in \mathbb{R}^{64}$, $\mathbf{e}_{g} \in \mathbb{R}^{16}$, $\mathbf{e}_{o} \in \mathbb{R}^{64}$, $\mathbf{e}_{m} \in \mathbb{R}^{16}$, and $\mathbf{e}_{a} \in \mathbb{R}^{16}$.

We implemented the Actor-Critic method \cite{peters2008natural} for each agent in our framework. This method includes an actor, which uses the policy $\pi$ with learning parameters $\theta^{\mu}$ to select an action in a given state, and a critic, which evaluates the chosen action using an action value function $Q$ with learning parameters $\theta^{w}$. In addition, we followed the implementation details of the Deep Deterministic Policy Gradient (DDPG) algorithm \cite{lillicrap2015continuous}. Each agent's actor network consists of two fully-connected Multi-Layer Perceptrons (MLP), each layer containing 128 neuron units. This configuration is also used for the agent's critic network. Adam \cite{kingma2014adam} is employed as the optimizer for learning the neural network parameters, with a learning rate of $10^{-4}$ for the actor and $10^{-3}$ for the critic. The discount factor $\gamma$ is set to 0.99, and the soft target update rate $\tau$ is set to $10^{-3}$. Furthermore, the batch size of the relay buffer $\mathcal{B}$ is 64.

We also designed a sparse reward function of an agent as follows:
\begin{equation}
    R(s^{t}_{i}) = \begin{cases}
        1 & \text{if } s^{t}_{i} = g_{i} \\
        -\lambda_{\text{stay}} & \text{if } \left( s^{t - 1}_{i} = s^{t}_{i}  \right) \land \left( s^{t}_{i} \neq g_{i} \right) \\
        (r_{\text{agg}})^{t}_{i} & \text{if } \left(s^{t - 1}_{i} \neq s^{t}_{i} \right) \\
        -1 & \text{otherwise} \\
    \end{cases}
\end{equation}
The reward value ranges between -1 and 1. An agent receives a reward of 1 if its position matches its goal. If the agent remains in a cell that is not its goal, it is penalized by $\lambda_{\text{stay}} \in (0, 1)$. In our experiments, we applied $\lambda_{\text{stay}} = 0.5$ to encourage an agent to keep moving. Moreover, to incentivize movement towards the goal, an agent receives a reward of $r_{\text{agg}}$ that is the same as Equation \ref{eq:reward}. Specifically, $r_{\text{ext}} = 1 - \Delta\left(s^{t}_{i}, g_{i}\right)$, where $\Delta$ is the geometric distance between $s^{t}_{i}$ and $g_{i}$, for each move, indicating that the closer the agent is to the goal, the higher the reward it receives. Inspired by \cite{skrynnik2024learn}, our experiments utilized the shortest path between $s^{t}_{i}$ and $g_{i}$ for $\Delta$ as follows: $\Delta\left(s^{t}_{i}, g_{i}\right) = \min\left(d(s^{t}_{i}, g_{i})\right)$. From our empirical analysis, we found that $\alpha \in [0.1, 0.5]$ in Equation \ref{eq:reward} tends to yield high outcomes, and hence, $\alpha = 0.1$ is the selected value for our experiments. Furthermore, the number of episodes and steps per episode are 100 and 300, respectively. Hence, we applied the time increment of 0.01 for $d_{t}$ in Equation \ref{eq:time-nov} for gradually decaying the value of agent's knowledge. The average reward of an agent at each episode is estimated as follows:
\begin{equation}
\begin{aligned}
    \text{AvgR} = \frac{1}{T_{i}}\sum_{t=1}^{T_{i}}{r^{t}_{i}} & \quad \text{where } T_{i} \leq \mathcal{T}
\end{aligned}
\end{equation}
where $T_{i}$ is the total number of steps taken by agent $i$ in one episode, and $\mathcal{T}$ is the maximum number of steps that an agent is allowed to take per episode. The overall performance of the system is then estimated as:
\begin{equation}
    R_{\text{overall}} = \frac{1}{M}{\sum_{i=1}^{M}{\frac{1}{N}\sum_{j=1}^{N}{\text{AvgR}_{j}}}}
    % \frac{1}{M}{\sum_{i=1}^{M}R_{i}}
\end{equation}
where $M$ is the number of episode, and $N$ is the number of agents. We aim to evaluate the performance of an agent in our framework based on both the average rewards and the number of steps an agent taken until reaching its goal. In terms of agent's coordination and knowledge aggregation, we set a threshold for $J$ in Equation \ref{eq:jac} as $J \leq 0.5$. This is designed to encourage the agent to learn from the substantial amount of novel knowledge shared by its peers. Additionally, we set $\beta = 0.1$ in Equations \ref{eq:agg_policy} and \ref{eq:agg_value} to prevent the agent's current knowledge from being overwhelmed by the influx of new knowledge.

In our experiments, we designed the following types of agents:
\begin{enumerate}
    \item \textbf{Independent Agents with DDPG ($A^{1}$)}: This type of agent follows the pure implementation of multi-agent DDPG (MADDPG) \cite{lowe2017multi}. However, agents are operated in a fully decentralized setting instead of adopting the framework of Centralized Training with Decentralized Execution (CTDE). Since this type of agent does not have time awareness, the intrinsic reward in Equation \ref{eq:reward} is always set to 0.
    \item \textbf{$A^{1}$ with Mental State ($A^{2}$)}: In comparison to $A^{1}$, an additional feature of this type of agent is the utilization of mental state per agent ($\mathcal{M}_{i}$). Note that $A^{2}$ still does not have time awareness.
    \item \textbf{$A^{2}$ with Time Awareness ($A^{3}$)}: $A^{3}$ is the extension of $A^{2}$ by having an additional feature of time awareness. However, $A^{3}$ is still an independent agent without the capability of communication and coordination.
    \item \textbf{$A^{3}$ with Communication and Coordination ($A^{4}$)}: $A^{4}$ extends $A^{3}$ by being equipped with the capability of communication and coordination. However, $A^{4}$ agents does not have goal awareness during the communication and coordination sessions. Hence, an agent is always an advisor of the other agent. Additionally, it shares its observations regardless the other agent's goal.
    \item \textbf{$A^{4}$ with Goal Awareness for Coordination ($A^{5}$)}: $A^{5}$ is the ultimate type of agent in our experiments. This type of agent have both time awareness and goal awareness for communication and coordination. Hence, $A^{5}$ is equipped with all features in this study.
\end{enumerate}
This design aims to evaluate the impact of each component that is integrated into an independent agent in the fully decentralized setting.

\subsection{Results and Discussion}
\subsubsection{Scenario 1}
Table \ref{tab:tab1} presents the experimental results for the first scenario, where all agents pursue the same goal. The integration of both mental state and time-awareness in independent agents within a fully decentralized setting ($A^{2}$ through $A^{5}$) generally yields better outcomes compared to $A^{1}$. Specifically, in the Base-Easy environment, $A^{5}$ outperforms the other agents, completing tasks with 5\% fewer steps on average. In the Base-Hard environment, not only does the performance of all agent types improve, but the number of steps taken is also reduced by 15\% compared to the Base-Easy environment. Notably, $A^{2}$ outperforms the others in this scenario, potentially due to dynamic obstacles creating pathways that allow faster goal achievement. Furthermore, $A^{4}$ and $A^{5}$ excel in the Large-Easy and Large-Hard environments, respectively, highlighting the importance of time-awareness for effective exploration in larger observation spaces. Additionally, to achieve higher outcomes when dealing with dynamic environments, time-aware agents must communicate and coordinate with each other. Interestingly, we observed that agents only reached their goals in a few episodes within the large environments. A potential solution is to increase the number of episodes and the maximum steps per episode. As these numbers increase, it is also crucial to select an appropriate value for $d_{t'}$ in Equation \ref{eq:time-nov}.

\begin{table}[!ht]
    \centering
    \begingroup
    \footnotesize
    \renewcommand{\arraystretch}{1.3}
    \begin{tabular}{|c|c|c|c|c|c|}
        \hline
        & \textbf{Base-Easy} & \textbf{Base-Hard} & \textbf{Large-Easy} & \textbf{Large-Hard} \\ \hline
        $A^{1}$ & 0.118 $\pm$ 0.06 & 0.176 $\pm$ 0.07 & 0.233 $\pm$ 0.04  & 0.243 $\pm$ 0.03 \\ \hline
        $A^{2}$ & 0.135 $\pm$ 0.06 & \textbf{0.207 $\pm$ 0.06} & 0.232 $\pm$ 0.04 & 0.225 $\pm$ 0.03 \\ \hline
        $A^{3}$ & 0.132 $\pm$ 0.05 & 0.198 $\pm$ 0.05 & \textbf{0.239 $\pm$ 0.04} & 0.236 $\pm$ 0.04 \\ \hline
        $A^{4}$ & 0.106 $\pm$ 0.06 & 0.168 $\pm$ 0.06 & 0.216 $\pm$ 0.04 & \textbf{0.237 $\pm$ 0.04} \\ \hline
        $A^{5}$ & \textbf{0.139 $\pm$ 0.06} & 0.191 $\pm$ 0.05 & 0.229 $\pm$ 0.04 & 0.235 $\pm$ 0.03 \\ \hline
    \end{tabular}
    \endgroup
    \caption{The overall performance ($R_{\text{overall}}$) of all agent types when pursuing a single goal across four different environments.}
    \label{tab:tab1}
\end{table}

\subsubsection{Scenario 2}
By comparing Table \ref{tab:tab2} with Table \ref{tab:tab1}, we observe that agents generally achieve higher rewards in the second scenario compared to the first. As illustrated in Table \ref{tab:tab2}, the overall performance of $A^{2}$ through $A^{5}$ continues to surpass that of $A^{1}$. Moreover, time-aware agents equipped with communication and coordination capabilities ($A^{4}$ and $A^{5}$) excel in three environments: Base-Easy, Base-Hard, and Large-Hard. Although $A^{5}$ does not outperform the independent agents in the Large-Easy environment, it is notable that $A^{5}$ tends to take fewer steps and reaches its goals in more episodes than the other types of agents.

\begin{table}[!ht]
    \centering
    \begingroup
    \footnotesize
    \renewcommand{\arraystretch}{1.3}
    \begin{tabular}{|c|c|c|c|c|c|}
        \hline
        & \textbf{Base-Easy} & \textbf{Base-Hard} & \textbf{Large-Easy} & \textbf{Large-Hard} \\ \hline
        $A^{1}$ & 0.134 $\pm$ 0.05 & 0.203 $\pm$ 0.05 & 0.249 $\pm$ 0.03  &  0.251 $\pm$ 0.03 \\ \hline
        $A^{2}$ & 0.144 $\pm$ 0.05 & 0.22 $\pm$ 0.05 & \textbf{0.242 $\pm$ 0.05} & 0.243 $\pm$ 0.04 \\ \hline
        $A^{3}$ & 0.112 $\pm$ 0.05 & 0.223 $\pm$ 0.05  & 0.242 $\pm$ 0.04 & \textbf{0.247 $\pm$0.04} \\ \hline
        $A^{4}$ & \textbf{0.163 $\pm$ 0.05} & 0.208 $\pm$ 0.05 & 0.224 $\pm$ 0.03 & 0.246 $\pm$ 0.04 \\ \hline
        $A^{5}$ & 0.144 $\pm$ 0.05 & \textbf{0.225 $\pm$ 0.05} & 0.232 $\pm$ 0.04 & 0.233 $\pm$ 0.03 \\ \hline
    \end{tabular}
    \endgroup
    \caption{The overall performance ($R_{\text{overall}}$) of all agent types when at least two agents pursue the same goal, while the remaining agents pursue a different goal across four distinct environments.}
    \label{tab:tab2}
\end{table}

\subsubsection{Ablation Study}
We conducted an ablation study to assess the contribution of each additional feature for independent agents in a fully decentralized environment. The first feature examined was the mental state of an agent ($A^{2}$), which generally enhances the performance of $A^{1}$ in the Base environments across both scenarios. However, this feature alone is insufficient for agents operating in the Large environments. To address this limitation, time awareness was introduced as an additional feature ($A^{3}$). The results in Tables \ref{tab:tab1} and \ref{tab:tab2} highlight the improvement of agents equipped with both mental state and time awareness compared to $A^{1}$. When communication and coordination were integrated into $A^{3}$, performance improvements were observed in three environments—Base-Easy, Base-Hard, and Large-Hard—across both scenarios. Furthermore, enhancing agent performance in Hard environments is crucial for managing dynamics in a fully decentralized setting. The introduction of goal-awareness also led to performance gains in the Base environments. Our ablation study demonstrates significant improvements in the performance of independent agents within a decentralized setting when equipped with mental state, time awareness, goal awareness, and a strategy that integrates communication and coordination.

\section{Conclusion and Future Work}
We proposed a novel Decentralized Muti-Agent Reinforcement Learning (Dec-MARL) framework that aims to address two key challenges such as exhaustive exploration and inefficient knowledge sharing among agent in the fully decentralized settings. Our framework introduces several innovative aspects: (i) the incorporation of an agent's mental state with time awareness; (ii) time-aware intrinsic rewards that motivate agents to explore novel states, potentially aiding in the achievement of their individual goals; (iii) the integration of communication and coordination; and (iv) the inclusion of goal awareness within this integration to facilitate efficient knowledge sharing. Experimental results demonstrate that our framework progressively enhances the performance of independent agents in fully decentralized 2D environments, where observation spaces may vary in size and obstacles can appear dynamically. Several potential directions for future work have emerged. Our experiments indicate that agents may require additional training time to achieve their goals in larger environments. Therefore, it is crucial to meticulously evaluate configurations related to time awareness to attain this. Additionally, while agents in our framework engage in peer-to-peer communication, they do not form any organizational structure despite having the same goal. Establishing an organization based on overlapping goals may accelerate and stabilize the exploration process, making it a promising area for further investigation.

\section*{Acknowledgment}
This work is supported as part of the Higher Degree Research (HDR) program at the Applied Artificial Intelligence Institute $(A^2I^2)$, Deakin University.

% \section*{References}

\bibliographystyle{ieeetr}
\bibliography{references}

\section*{Appendix 1 - Description of Datasets}
\subsection{The Base Environment}

\begin{figure}[h]
    \centering
    \includegraphics[width=0.8\linewidth]{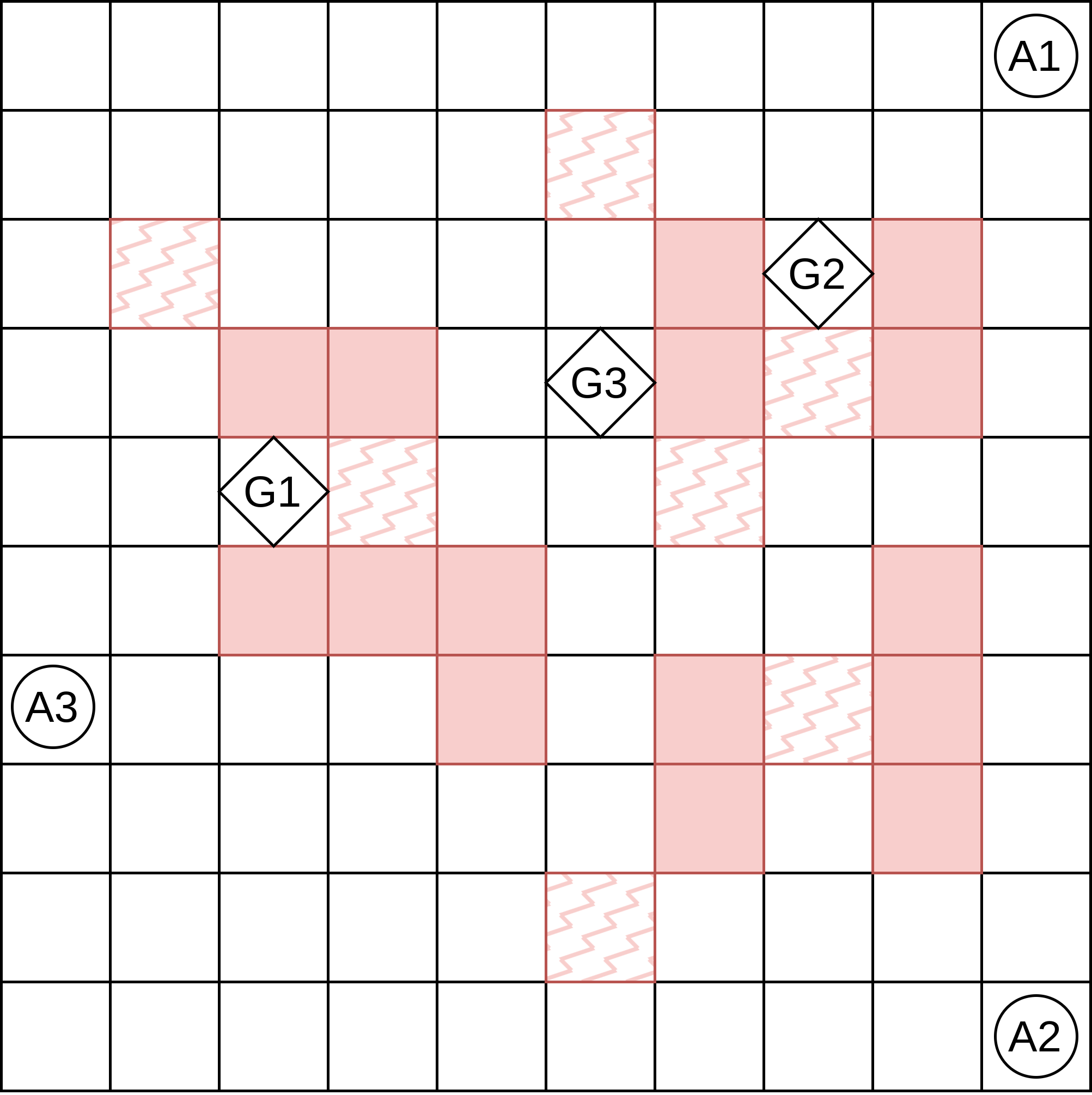}
    \caption{An illustration of the Base environment}
    \label{fig:base}
\end{figure}

Figure \ref{fig:base} shows the Base environment mentioned in the paper. A circle represents an agent, a diamond represents a goal of agents, a box filled by the red color represents an obstacle, and a box filled by zigzag lines represents an obstacle that is dynamically occurring. Furthermore, there are two settings such as: (i) all agents pursuing a single goal (i.e., G3); and (ii) Agents 1 and 2 pursuing Goal 1 and Agent 3 pursuing Goal 2.

\subsection{The Large Environment}

\begin{figure*}[tp]
    \centering
    \includegraphics[width=0.9\linewidth]{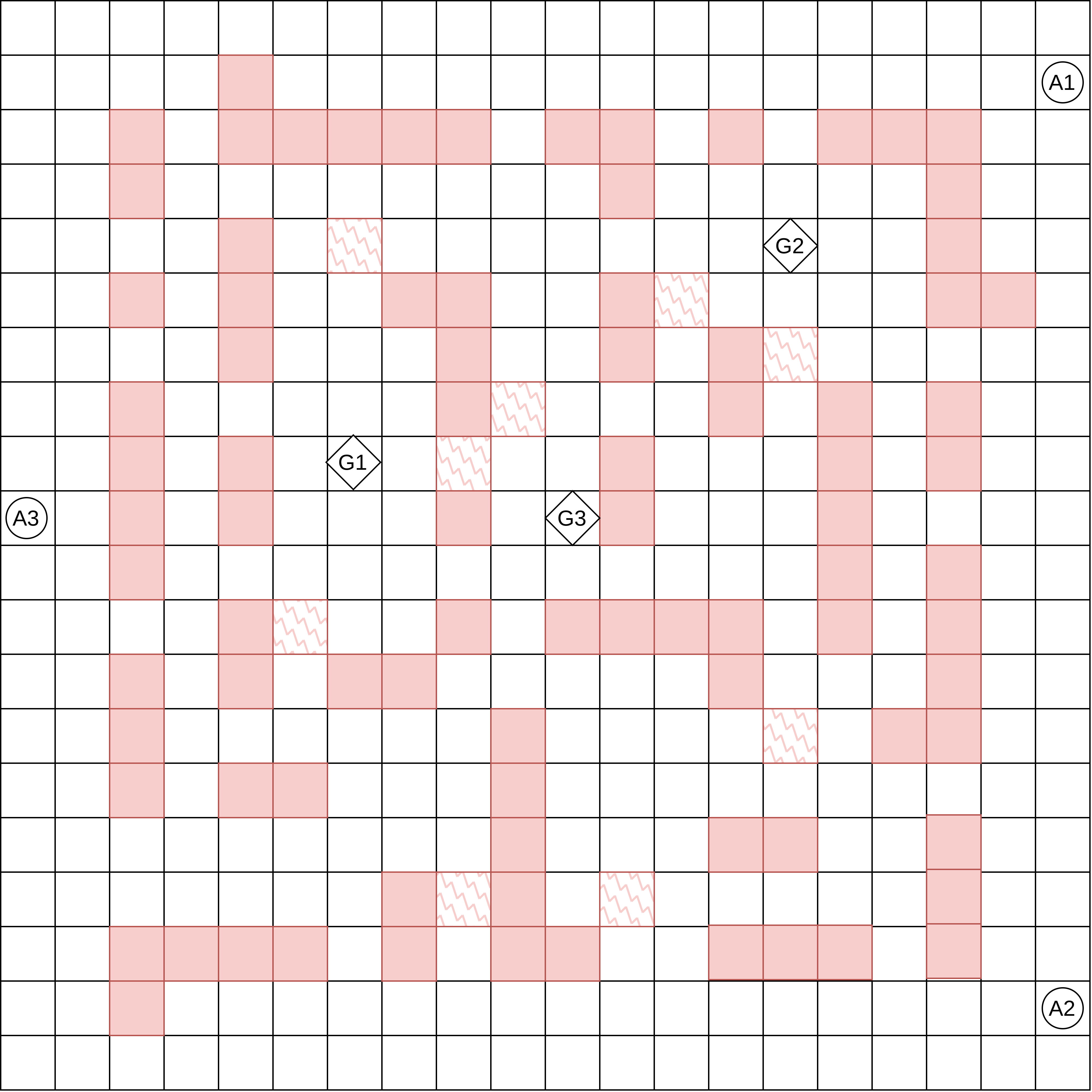}
    \caption{An illustration of the Large environment}
    \label{fig:large}
\end{figure*}

Figure \ref{fig:large} shows the Large environment mentioned in the paper. A circle represents an agent, a diamond represents a goal of agents, a box filled by the red color represents an obstacle, and a box filled by zigzag lines represents an obstacle that is dynamically occurring. Furthermore, there are two settings such as: (i) all agents pursuing a single goal (i.e., G3); and (ii) Agents 1 and 2 pursuing Goal 1 and Agent 3 pursuing Goal 2.

\end{document}